\documentclass[nofootinbib,twocolumn,aps,amsmath,amssymb]{revtex4}
\usepackage{graphics,bm}
\usepackage{epsfig}
\usepackage{graphicx}

\newcommand{\beq}{\begin{equation}}
\newcommand{\eeq}{\end{equation}}
\newcommand{\bqa}{\begin{eqnarray}}
\newcommand{\eqa}{\end{eqnarray}}

\newcommand{\unity}{1\hspace{-1.3mm}1}

\def\slashchar#1{\setbox0=\hbox{$#1$}     		
   \dimen0=\wd0                                 	
   \setbox1=\hbox{/} \dimen1=\wd1               	
   \ifdim\dimen0>\dimen1                        	
      \rlap{\hbox to \dimen0{\hfil/\hfil}}      	
      #1                                        	
   \else                                        	
      \rlap{\hbox to \dimen1{\hfil$#1$\hfil}}   	
      /                                         	
   \fi}      

\begin{document}

\parindent=20pt
\pagestyle{plain}
\font\tenrm=cmr10

\author{Harmen J. Warringa and Dani\"el Boer} \affiliation{Department
of Physics and Astronomy, Vrije Universiteit, De Boelelaan 1081, 1081
HV Amsterdam, The Netherlands} 
\email{harmen@nat.vu.nl, dboer@nat.vu.nl} 

\title{Color superconductivity
vs.\ pseudoscalar condensation in a three-flavor NJL model} 
\author{Jens O. Andersen}
\affiliation{Nordita, Blegdamsvej 17, DK-2100 Copenhagen {\O},
Denmark} \email{jensoa@nordita.dk} \ \date{\today}

\begin{abstract}
We calculate numerically the phase diagram of the three-flavor
Nambu--Jona-Lasinio model at zero and finite temperature as a function
of the up, down, and strange quark chemical potentials.  
We focus on the competition between
pseudoscalar condensation and color superconductivity.  We find that
the two types of phases are separated by first-order transitions.
\end{abstract}

\maketitle

\section{Introduction}
In recent years a huge effort has been made in order to understand the
phase diagram of strongly interacting matter, both experimentally as
well as theoretically. Collisions at the Relativistic Heavy-Ion
Collider (RHIC) at BNL and the Large Hadron Collider (LHC) at CERN
allow the experimental study of hadronic matter at energy densities
exceeding that required to create a quark-gluon plasma. The energy and
baryon densities of these experiments correspond to temperatures up to
200 MeV and a baryon chemical potential in the range 0-600 MeV.

At zero baryon chemical potential, lattice calculations suggest the
existence of a transition from ordinary nuclear matter to a
quark-gluon plasma at temperatures of around 160 MeV (see for example
Ref.\ {\cite{Karsch2000}}). While there has been progress in
generalizing lattice calculations to finite baryon chemical potential
\cite{Fodor2002b, deForcrand2002, Allton2002, DElia2003}, highly
nontrivial problems remain to be solved, and in practice one is
restricted to small baryon chemical potentials.

Since one cannot apply lattice gauge theories to the region of the
phase diagram where the baryon chemical potential is large, one must
employ other methods. One possibility is perturbative QCD. 
Asymptotic 
freedom guarantees that this method can be applied at asymptotically
large chemical potentials. If the densities are too small to
use perturbative QCD, one may apply effective theories. Models such as
the instanton liquid model, random matrix models, and
Nambu--Jona-Lasinio (NJL) models have been applied to the study of the
QCD phase diagram at finite temperature and densities. Despite their
shortcomings, it is expected that such models do describe the
qualitative features of the QCD phase diagram, in regions not
accessible to perturbative or lattice QCD. For example, the existence
of a critical point at finite baryon chemical potential has been predicted 
using effective models~\cite{Halasz1998, Berges1998}. Furthermore, it
is expected that quark matter at high baryon chemical potentials and
low temperatures can be in a two-flavor color-superconducting phase
(2SC) \cite{Bailin1984, Alford1997, Rapp1997} or a color-flavor locked
(CFL) phase \cite{Alford1999, Alford1999q}. In these phases gaps on the order of
100 MeV arise.

Many results have been obtained for equal up, down and strange quark
chemical potentials. However, this may not be directly relevant for
heavy-ion collisions or compact stars. For example, to enforce
electric and color neutrality and weak equilibrium in compact stars
different flavor and color chemical potentials have been
introduced~\cite{Alford2002, Steiner2002}.  This gives rise to a more
complicated phase diagram in which one also finds new 2SC-like
\cite{Neumann2002} and gapless superconducting phases
\cite{Alford1999b, Shovkovy2003, Alford2004a, Alford2004c, Ruster2004, Abuki2004, Ruster2005,
Blaschke2005}. Similarly, in heavy-ion collisions, a difference
between the quark chemical potentials arises if the number densities
of the different quark flavors are not the same. This difference can
cause interesting observable effects such as two critical
endpoints~\cite{Klein2003, Toublan2003}. However, instanton induced
interactions tend to suppress this effect \cite{Frank2003}.

In addition to a more complicated structure of the superconducting
phases, different chemical potentials can also trigger pseudoscalar
condensation \cite{Son2001, Kogut2001, Barducci2003}. This has been
confirmed on the lattice at zero baryon but finite isospin chemical
potential \cite{Kogut2002}. Depending on the flavors involved, the
charged pion, neutral or charged kaon fields may acquire a vacuum
expectation value. Pseudoscalar condensation in the two-flavor NJL
model has been studied in Refs.\ \cite{Toublan2003, Barducci2004}, as
a function of the different chemical potentials at zero and finite
temperature.  An extension to three flavors was carried out in Ref.\
\cite{Barducci2005}.

In this paper, we discuss the phase diagram of the three-flavor NJL
model as a function of the different chemical potentials including
both pseudoscalar condensation and color superconductivity. We did not
apply electric or color neutrality conditions. At zero temperature,
pseudoscalar condensation is possible if $\vert \mu_u - \mu_d \vert >
m_\pi$ \cite{Son2001} or if $\vert \mu_{u,d} - \mu_s \vert > m_K$
\cite{Kogut2001}. On the other hand, color superconducting phases
occur if the chemical potentials are large and approximately
equal. Therefore, one can imagine scenarios where for example $\mu_u
\approx \mu_d$ (the Fermi surfaces of the $u$ and $d$ quark should be
sufficiently close for Cooper pairing to occur) and $\mu_u \approx -
\mu_s$ (the Fermi surfaces of the $u$ and $\bar{s}$ should be
sufficiently close for kaon condensation to occur), with $\vert
\mu_{u,d} - \mu_s \vert > m_K$. In such a case a 2SC phase is
competing against a phase in which kaons condense.  From our
calculations it follows that a coexistence phase of pseudoscalar condensation and
color superconductivity does not occur for the
parameters chosen and that these phases are separated by a first-order
transition. However, we do not exclude that other choices of
parameters may lead to such a coexistence phase, just as coexistence of
color superconductivity and chiral symmetry breaking may occur in the
NJL model for specific ranges of parameters \cite{Blaschke2003}. Here we refer
to a coexistence phase to mean a phase in which two condensates are nonzero
simultaneously.

In this work we study pseudoscalar condensation in the quark
anti-quark channel. We have not taken into account the pseudoscalar
diquark interaction.  This interaction is suppressed relative to the
scalar diquark interaction due to instantons~\cite{Rapp1997}. However
in absence of instanton interactions, if one neutralizes the bulk
matter with respect to color and electric charges it is possible to
have pseudoscalar diquark condensation with rather large gaps
\cite{Buballa2004b}. Pseudoscalar diquark condensation in the NJL
model is similar \cite{Buballa2004b} to pseudoscalar condensation in
the CFL phase studied with effective chiral models in Refs.\
\cite{Casalbuoni1999,Son1999,Casalbuoni2000, Schafer2000,
Bedaque2002a, Kaplan2002, Forbes2004}.

We emphasize that in this paper we do not impose electric or color neutrality
conditions, but that this would qualitatively affect the phase structure, 
leading for example to the observation that in a macroscopic volume of quark
matter the 2SC phase is energetically disfavored~\cite{Alford2002,
Steiner2002}. Therefore, our
results do not address the qualitative features of the QCD phase structure 
with neutrality constraints imposed.

The paper is organized as follows. In Sec.\ II, we briefly describe
the NJL model and the choice of parameters.  In Sec.\ III, we discuss the
calculations and some of its technical aspects. In Sec.\ IV, we
present our results, and in Sec.\ V we conclude.

\section{The NJL model}
In the NJL model \cite{Nambu1961}, one treats the interaction between
the quarks as a point-like quark color current-current
interaction. This naive approximation to QCD works very well in
explaining various low-energy observables such as hadron masses
\cite{Klevansky1992}.  By applying several Fierz transformations to
the current-current interaction (see for example Ref.\
\cite{Buballa2004a}) and including only terms which give rise to
attractive $qq$ and $\bar{q}q$ channels, one obtains the following
Lagrangian density
\begin{equation}
  \mathcal{L} = \bar \psi \left(i \gamma^\mu \partial_\mu - M_0
  + \mu \gamma_0 \right)\psi 
  + \mathcal{L}_{\bar q q} + \mathcal{L}_{qq}
 \label{eq:lagrnjl} \;,
\end{equation}
where the quark-antiquark term, $\mathcal{L}_{\bar q q}$, and the
diquark interaction term, $\mathcal{L}_{q q}$, are defined below.  We
have suppressed the color, flavor, and Dirac indices of the fermion
fields $\psi$ for notational simplicity.  The mass matrix $M_0$ is
diagonal and contains the bare quark masses $m_{0u}$, $m_{0d}$ and
$m_{0s}$. The matrix $\mu$ is also diagonal and contains the quark
chemical potentials $\mu_u$, $\mu_d$ and $\mu_s$.  We use the metric
$g^{\mu \nu} = \mathrm{diag}(+\,-\,-\,-)$ and the standard representation
for the $\gamma$-matrices.  The quark-antiquark interaction part of
the Lagrangian density is
\begin{equation}
 \mathcal{L}_{\bar q q} 
  = G \left[ \left(\bar \psi \lambda_a \psi \right)^2 + 
 \left(\bar \psi \lambda_a i \gamma_5 \psi \right)^2 \right] \;.
\end{equation}
The matrices $\lambda_a$ are the $9$ generators of $\mathrm{U}(3)$ and
act in flavor space. They  
are normalized as $\mathrm{Tr}\, \lambda_a
\lambda_b = 2 \delta_{a b}$.  
The diquark interaction term of the Lagrangian density is given by
\begin{equation}
 \mathcal{L}_{qq} =  
  \frac{3}{4} G \left(\bar \psi t_{A} \lambda_{B} C i \gamma_5 \bar \psi^T \right)
\left(\psi^T  t_{A} \lambda_{B} C i \gamma_5 \psi \right)
 \;,
\end{equation}
where $A, B \in \{2, 5, 7\}$ since only the interaction in the color
and flavor antisymmetric triplet channel is attractive. The matrices
$t_a$ are the generators of $\mathrm{U}(3)$ and act in color space.
Their normalization is $\mathrm{Tr}\, t_a t_b = 2 \delta_{a b}$.  To
remind the reader, the antisymmetric flavor matrices $\lambda_2$,
$\lambda_5$ and $\lambda_7$ couple up to down, up to strange and down
to strange quarks, respectively.  The charge conjugate of a field
$\psi$ is denoted by $\psi_c = C \bar \psi^T$ where $C = i \gamma_0
\gamma_2$.  The coupling strength $3 G /4$ of the diquark interaction
is fixed by the Fierz transformation. However, some authors discuss the NJL model
with a different diquark coupling constant (see for example Ref.\
\cite{Ruster2005} for a comparison).

The results that will be presented below are obtained with the
following choice of parameters
\begin{eqnarray}
 m_{0u} = m_{0d} = 5.5\;\mathrm{MeV} \;,\;\; m_{0s} =
112\;\mathrm{MeV}
 \nonumber \\
 G = 2.319 / \Lambda^2 \;, \;\;\Lambda = 602.3\;\mathrm{MeV} \;.
\end{eqnarray}
This choice of parameters gives rise to constituent quark masses
$M_u=M_d=368$ MeV and $M_s=550$ MeV \cite{Buballa2004a}. 

\section{The effective potential}
To obtain the phase diagram of the NJL model, we first introduce 18
real condensates $\alpha_a$ and $\beta_a$, and 9 complex condensates
$\Delta_{AB}$ as follows
\begin{eqnarray}
  \alpha_a  &=& - 2 G \left < \bar \psi \lambda_a \psi 
\right>\;, \\
  \beta_a  &=& - 2 G \left < \bar \psi \lambda_a i
  \gamma_5 \psi
  \right> \;, \\
  \Delta_{AB} &=& \frac{3}{2}G \left < \psi^T t_A \lambda_B C \gamma_5
\psi \right> \;.
\end{eqnarray}
We will assume that all condensates are space-time independent. The
crystalline Larkin-Ovchinnikov-Fulde-Ferrell (LOFF) phase
\cite{Alford2001b} will not be considered here. The next step is
to apply a Hubbard-Stratonovich transformation to eliminate the
four-point quark interactions to make the Lagrangian quadratic in the
quark fields. Introducing a two-component Nambu-Gorkov field $\Psi^T =
\left( \psi^T\!,\;\psi_c^T \right) / \sqrt{2}$ allows for
straightforward integration over the quark fields. After going to
imaginary time, the thermal effective potential ${\cal V}$ in the mean-field 
approximation reads
\begin{multline}
  \mathcal{V} = \frac{\alpha_a^2 + \beta_a^2}{4G}
+ \frac{\left \vert \Delta_{AB}\right \vert^2 }{3 G} \\
-\frac{T}{2} \sum_{p_0 = (2n + 1)\pi T} \int \frac{\mathrm{d}^3 p}{
\left( 2\pi \right)^3} \log \mathrm{det} K \;,
\label{eq:effpotential}
\end{multline}
where $K$ is a $72 \times 72$ matrix
\begin{equation}
K = 
\left(
\begin{array}{cc}
\unity_c \otimes \mathcal{D}_1 
&
\Delta_{AB}\, t_A \otimes \lambda_B \otimes \gamma_5 
\\
-\Delta^*_{AB}\,  t_A \otimes \lambda_B \otimes \gamma_5 
&
\unity_c \otimes \mathcal{D}_2
\end{array} \right)\;,
\end{equation}
and 
\begin{eqnarray}
  \mathcal{D}_1 &=& \unity_f \otimes (i \gamma_0 p_0 +
\gamma_i p_i) 
-  \mu \otimes \gamma_0 \nonumber \\
& &
  - ( M_0 +\alpha_a \lambda_a ) \otimes \unity_d 
- \beta_a \lambda_a \otimes i \gamma_5 
\;,
\\
  \mathcal{D}_2 &=& \unity_f \otimes (i \gamma_0 p_0 + \gamma_i
p_i) 
 +  \mu \otimes \gamma_0
  \nonumber  \\
& &
 - ( M_0 + \alpha_a \lambda_a^T)\otimes \unity_d  
 - \beta_a \lambda_a^T \otimes i \gamma_5
\;.
\end{eqnarray}
The matrix $\unity$ is the identity matrix in color ($c$), flavor ($f$),
or Dirac ($d$) space.

The values of the condensates and the phase diagram are determined by
minimizing the effective potential ${\cal V}$ with respect to the
condensates. To make the minimization procedure easier, one can take
advantage of the fact that certain condensates must vanish. Firstly,
application of QCD inequalities \cite{Weingarten1983, Son2001} shows
that in QCD at zero chemical potential chiral symmetry breaking cannot be
driven by parity-violating condensates of the type $\left < \bar \psi
i \gamma_5 \psi \right>$. Outside the phase in which diquarks condense, we
found numerically that this is also correct in the NJL model at finite chemical potentials.
Therefore, $\beta_0$, $\beta_3$ and
$\beta_8$ are put to zero. Secondly, although
perturbative one-gluon exchange cannot distinguish between $\beta_k$
and $\alpha_k$ condensation with $k \in \{1,2,4,5,6,7\}$, pseudoscalar
condensation is favored due to the instanton interaction
\cite{Son2001}. We therefore set all $\alpha_k$'s with $k \in
\{1,2,4,5,6,7\}$ to zero. We found numerically that this is correct,
despite the fact that the model we consider does not include instanton
interactions.

One can further simplify the minimization procedure by using the
symmetries of the NJL model. In absence of quark masses and chemical
potentials, the Lagrangian density has a global
$\mathrm{SU(3)}_c\times\mathrm{U(3)}_{V}\times\mathrm{U(3)}_{A}$
symmetry. Due to the non-vanishing quark masses, the symmetry is
broken down to $\mathrm{SU(3)}_c \times \mathrm{U(3)}_{V}$. Since we
consider different quark masses and finite chemical potentials, the
symmetry of the Lagrangian density is further reduced to
$\mathrm{SU(3)}_c\times\mathrm{U(1)}_u\times\mathrm{U(1)}_d
\times\mathrm{U(1)}_s $. The vacuum manifold is invariant under the
same transformations as the Lagrangian density, so applying a
$\mathrm{U}(1)$-flavor transformation to all condensates leaves the
free energy invariant. Therefore, using the $\mathrm{U}(1)$-flavor
transformations one can choose the pseudoscalars to condense in the
$\beta_2$, $\beta_5$, and $\beta_7$ channels, and set $\beta_1$,
$\beta_4$ and $\beta_6$ to zero. The phase in which $\beta_2$,
$\beta_5$ or $\beta_7$ is non-vanishing is called the $\pi^+/\pi^-$,
$K^0 / \bar K^0$ or $K^+/K^-$ condensed phase, respectively.

Because of the global $\mathrm{SU}(3)_c$ symmetry, one can also rotate
away several diquark condensates. Without loss of generality, we can
minimize with respect to $\Delta_{22}$, $\Delta_{25}$, $\Delta_{55}$,
$\Delta_{27}$, $ \Delta_{57}$ and $\Delta_{77}$. In principle, all six
diquark condensates can have a phase. It is always possible to remove
two of them by using the two diagonal $\mathrm{SU}(3)_c$
transformations. As long as there is no pseudoscalar condensation, one
can use the $\mathrm{U}(1)$-flavor symmetries to rotate away three
other phases. As a result either $\Delta_{25}, \Delta_{55},
\Delta_{27}$, or $\Delta_{57}$ has a phase \cite{Buballa2004a}.
However, this reduction is not completely possible if pseudoscalar
condensation occurs. By choosing the pseudoscalars to condense in the
$\beta_2$, $\beta_5$ and $\beta_7$ channels, one breaks the
$\mathrm{U}(1)$-flavor symmetry. Hence if pseudoscalar condensation
arises in one channel, one can in general rotate away one phase less in the diquark
sector. If it occurs in more channels, two phases
less can be rotated away.
However, numerically we find that allowing for a
complex phase leads to diquark condensation only in the $\Delta_{22}$,
$\Delta_{55}$ and the $\Delta_{77}$ channels. The $\Delta_{25}$,
$\Delta_{27}$, and $\Delta_{57}$ diquark condensates do not arise or
can be rotated away. Moreover, we find that pseudoscalar condensation in 
the quark-antiquark channel
does not coexist with color superconductivity, such that one can always 
take the diquark condensates to be real.

The different possible color-superconducting phases are
named as follows \cite{Ruster2005}
\begin{eqnarray}
 \Delta_{22}\neq0,\,\;\Delta_{55}\neq0,\,\Delta_{77} \neq 0 & & \mathrm{CFL} \;, 
  \nonumber \\
 \Delta_{77} =0,\;\; \Delta_{22}\neq0,\,\Delta_{55} \neq 0 & & \mathrm{uSC} \;,
  \nonumber \\
 \Delta_{55} = 0,\;\;\Delta_{22}\neq0,\,\Delta_{77} \neq 0 & & \mathrm{dSC} \;,
  \nonumber \\
 \Delta_{22} = 0,\;\;\Delta_{55}\neq0,\,\Delta_{77} \neq 0 & & \mathrm{sSC} \;,
  \nonumber \\
 \Delta_{22} \neq 0, \;\; \Delta_{55}=0,\, \Delta_{77} = 0 & & \mathrm{2SC} \;,
  \nonumber \\
 \Delta_{55} \neq 0, \;\; \Delta_{22}=0,\, \Delta_{77} = 0 & & \mathrm{2SCus} \;,
  \nonumber \\
 \Delta_{77} \neq 0, \;\; \Delta_{22}=0,\, \Delta_{55} = 0 & & \mathrm{2SCds} \;.
\end{eqnarray}

To calculate the effective potential in an efficient way, one can multiply
the matrix $K$ with $\mathrm{diag}(\unity_c \otimes \unity_f \otimes
\gamma_0, \unity_c \otimes \unity_f \otimes \gamma_0 )$ which leaves
the determinant invariant.  In this way, one obtains a new matrix $K'$
with $i p_0$'s on the diagonal.  By determining the eigenvalues of the
matrix $K'$ with $p_0 = 0$, one can reconstruct the determinant
of $K$ for all values of $p_0$ which is namely
$\prod_{i=1}^{72}(\lambda_i - ip_0)$. After summing over Matsubara
frequencies, one finds
\begin{equation}
 T \!\!\!\!\!\! \sum_{p_0 = (2 n+1) \pi T} \!\!\!\!\!\! \log
  \mathrm{det} K = \sum_{i=1}^{72} \left[\frac{\lambda_i}{2} + T
  \log\left(1 + e^{-\lambda_i / T} \right) \right] \;.
\end{equation}
All that remains in order to determine the effective potential is to
integrate over three-momentum $p$ up to an ultraviolet cutoff $\Lambda$. 

The speed of the calculation of the effective potential depends
heavily on how fast one can compute the eigenvalues. There are several
ways to speed up the calculation. Firstly, the determinant of $K$ does
not depend on the direction of $\vec p$. Therefore, one can choose 
$\vec p$
to lie in the $z$-direction.  Together with the choice of the
non-vanishing condensates mentioned above, this implies that
$K'(p_0=0)$ becomes a real symmetric matrix, which simplifies the
calculation of the eigenvalues. Secondly, one can interchange rows and
columns of $K'$ without changing its determinant. By doing so, one can
bring $K'$ in a block-diagonal form. One can then determine the
eigenvalues of the blocks separately which is significantly faster
since the time needed to compute eigenvalues numerically scales
cubically with the dimension of the matrix. In the most general case
with diquark condensation, one can always reduce the problem to two
$36 \times 36$ matrices. Moreover, if there is no diquark
condensation, but only pseudoscalar condensation, the problem can be
further reduced to computing the eigenvalues of two $6 \times 6$ matrices.

We determined the eigenvalues using LAPACK routines \cite{Lapack1999}.
After numerical integration over three-momentum $p$ up to the
cutoff, the condensates were determined by minimizing the effective
potential using MINUIT \cite{James1975}. To be certain that the
minimization procedure did not end up in a local minimum, we always
checked the continuity of the minimized effective potential as a
function of chemical potentials and/or temperature.

\section{Phase diagrams}
In this section, we present our results for the phase diagrams of the
NJL model with $u$, $d$, and $s$ quarks. We plot the phase diagrams
as a function of the chemical potentials and temperature. To determine
the locations of the phase boundaries, we examine the behavior of the
condensates. If a condensate jumps discontinuously
the transition is first-order, and this is indicated by a solid line.
If its derivative has a
discontinuity, the transition is second order, and this is indicated
by a dotted line.
If a condensate changes rapidly in a narrow range without
vanishing there is a smooth cross-over, and this is indicated by
a dashed-dotted line at the point were the condensate varies maximally. 

\subsection{$\mu_s=0$ and $T=0$}

\begin{figure}[htb]
\includegraphics{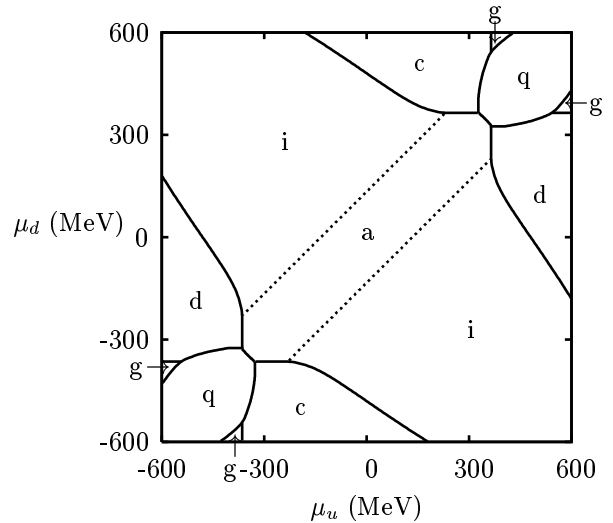}
\caption{Phase diagram for $\mu_s = 0$ and $T=0$ as a function of
$\mu_u$ and $\mu_d$. First and second-order transitions are indicated
by solid and dotted lines, respectively. The
letters denote the different phases, where 
a: $\bar u u$ + $\bar d d$ + $\bar s s$, 
c: $\bar u u$ + $\bar s s$, 
d: $\bar d d$ + $\bar s s$,
g: $\bar s s$, 
i: $\pi^+ / \pi^-$ + $\bar s s$ and
q: 2SC + $\bar s s$.}
\label{fig:muumudt0}
\end{figure}

In Fig.\ \ref{fig:muumudt0}, we display the phase diagram of the NJL
model for $\mu_s=0$ and $T=0$ as a function of $\mu_u$ and
$\mu_d$. Outside the 2SC phase (q), our results agree qualitatively with
the two-flavor calculations of Ref.\ \cite{Barducci2004} (see their
Fig.\ 1), where color-superconducting phases were not taken into account. 
Moreover, in Ref.\ \cite{Barducci2004} different parameters were used, in addition
to a form factor which mimics asymptotic freedom.

One can clearly see that the phase diagram is symmetric under
reflection in the origin. This is because the free energy is invariant
under the transformation $(\mu_u, \mu_d, \mu_s) \rightarrow (-\mu_u,
-\mu_d, -\mu_s)$ from the symmetry between particles and
anti-particles. Fig.\ \ref{fig:muumudt0} is also symmetric under
interchange of $u$ and $d$, because of the choice of equal up and down
quark masses. This gives rise to the symmetry of the phase boundaries
with respect to the diagonals.

In general, horizontal and vertical lines in the phase diagrams arise
if the pairing of one type of quark is not changed after a
transition. In this case, the location of the phase boundary is
determined by the properties of other quarks. Therefore, changing the
chemical potential of the unchanged quark species cannot have a big
influence on the location of the phase boundary. This results in the
horizontal and vertical lines. For $T=0$, one always finds these lines
near the values of the constituent quark masses, i.e.\ $\mu_u \approx
M_u$, $\mu_d \approx M_d$ and $\mu_s \approx M_s$ (see for example Ref.\
\cite{Buballa2004a}). The diagonal lines arise
because at $T=0$ pion condensation can occur if $\vert \mu_u - \mu_d
\vert > m_\pi = 138\;\mathrm{MeV}$ \cite{Son2001}.

The diagram shows that if the chemical potentials are different, the
transition to the color-superconducting phase (q) remains first order as
was concluded in Ref.\ \cite{Bedaque2002}.  Moreover, one can see from
Fig.\ \ref{fig:muumudt0} that if $\mu_u \neq \mu_d$ it is possible to
go through two first-order transitions before entering the 2SC phase (q)
(similar to the situation discussed in Ref.\ \cite{Toublan2003}
without color superconductivity). We observe that to have such a
scenario at zero temperature, a minimum difference between $\mu_u$ and
$\mu_d$ is required. In the present case this is about 35 MeV.  Pion
condensation (i) and the 2SC phase (q) are in this diagram separated by two
phase transitions in contrast to the estimated $(\mu_B,\mu_I)$ phase
diagram of Ref.\ \cite{He2005}.

\subsection{$\mu_d = 0$ and $T=0$}
\begin{figure}[htb]
\includegraphics{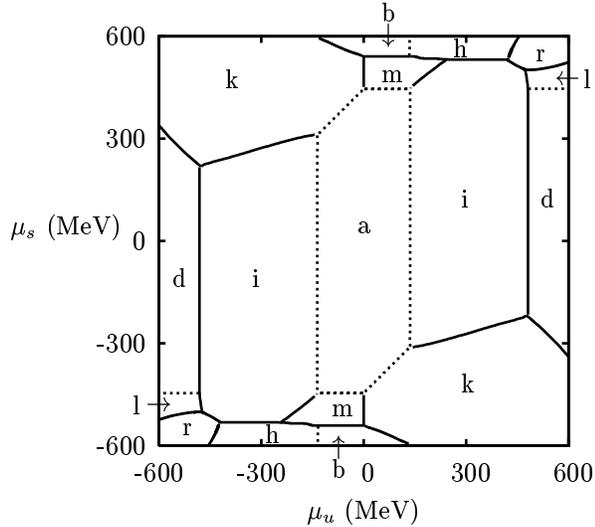}
\caption{Phase diagram for $\mu_d = 0$ and $T = 0$ as a function of
$\mu_u$ and $\mu_s$. First and second-order transitions are indicated
by solid and dotted lines, respectively. The letters denote the
different phases, where a: $\bar u u$ + $\bar d d$ + $\bar s s$, 
b: $\bar u u$ + $\bar d d$, 
d: $\bar d d$ + $\bar s s$, 
h: $\pi^+ / \pi^-$, i: $\pi^+ / \pi^-$ + $\bar s s$, 
k: $K^+/ K^-$ + $\bar d d$,
l: $K^0 / \bar K^0$, 
m: $K^0/\bar K^0$ + $\bar u u$
and r: 2SCus + $\bar d d$.}
\label{fig:muumust0} 
\end{figure}

In Fig.\ \ref{fig:muumust0} we display the phase diagram for $\mu_d =
0$ and $T=0$ as a function of $\mu_u$ and $\mu_s$. Since the up and
down quark masses are much smaller than the strange quark mass, this
diagram is very different from Fig. \ref{fig:muumudt0}. Besides the
possibility of pion condensation in (h) and (i), phases in which
the charged kaon (k) and the neutral kaon condense (l)/(m) arise. The
lines separating the charged kaon phase (k) from the chirally broken
phase (a) are diagonal because at $T=0$ kaon condensation can occur if
$\vert \mu_s - \mu_{u, d} \vert > m_K = 450\;\mathrm{MeV}$
\cite{Kogut2001} (the chosen parameter set gives rise to a somewhat low kaon
mass, but this is not relevant for the qualitative features of the phase 
diagram). 
In (r) we find the 2SCus phase. This phase is surrounded by
phases in which the pions (h)/(i) and the neutral kaons (l)
condense. We find that one passes a first-order transition when going
from the pion and neutral kaon condensed to the 2SCus phase.

\subsection{$\mu_u \approx \mu_d$}

\begin{figure}[htb]
\includegraphics{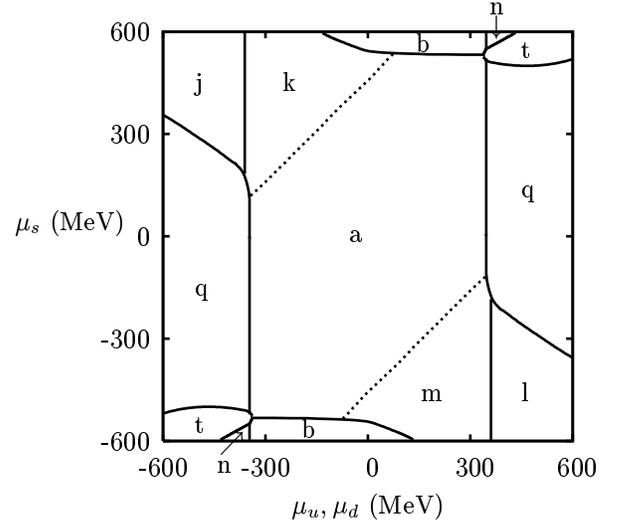}
\caption{Phase diagram for $T = 0$ as a function of $\mu_u = \mu_d +
\epsilon $ and $\mu_s$. First and second-order transitions are
indicated by solid and dotted lines, respectively.  The letters denote
the different phases, where 
a: $\bar u u$ + $\bar d d$ + $\bar s s$,
b: $\bar u u$ + $\bar d d$, 
j: $K^+ / K^-$,  
k: $K^+/ K^-$ + $\bar d d$,
l: $K^0 / \bar K^0$, 
m: $K^0/\bar K^0$ + $\bar u u$,
n: 2SC, 
q: 2SC + $\bar s s$ and 
t: CFL.
}
\label{fig:muudmust0} 
\end{figure}

In Fig.\ \ref{fig:muudmust0}, we display the phase diagram for $T=0$
as a function of the up and down quark chemical potential and the
strange quark chemical potential. We have chosen $\mu_u = \mu_d +
\epsilon$ where $\epsilon$ is a very small positive
number. This $\epsilon$ is necessary because when $\epsilon = 0$ one
is just at a first-order phase boundary between the phase in which the charged kaons condense
 (k) and the one in which the neutral kaons condense (m), 
as can be seen from Fig.\ \ref{fig:muumust0}. This nonzero value of
$\epsilon$ gives rise to a small asymmetry in the phase diagram. If
$\epsilon$ is chosen negative, the phases in which the neutral (l)/(m) and the charged (j)/(k) kaon condenses are
interchanged.

Apart from the additional 2SC (n)/(q) and CFL (t) phases, our results agree
qualitatively with the three-flavor calculations 
of Ref.\ \cite{Barducci2005} (see their Fig.\ 7).

\begin{figure}[htb]
\includegraphics{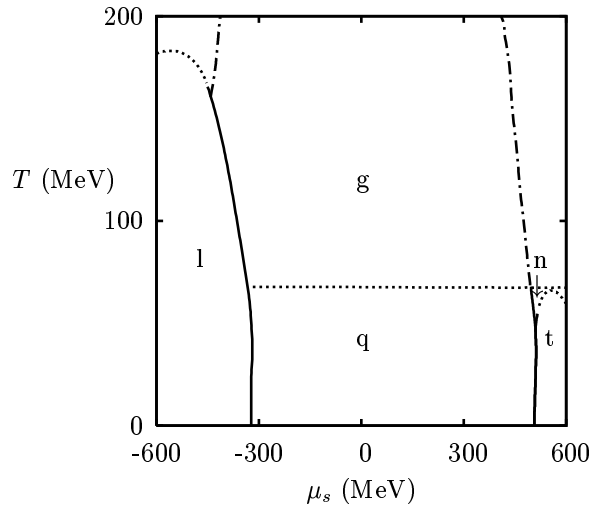}
\caption{Phase diagram as a
function of $\mu_s$ and $T$, for fixed 
$\mu_u = \mu_d = 550\;\mathrm{MeV}$. 
First and second-order 
transitions are indicated
by solid and dotted lines respectively, while
cross-overs are indicated by
dashed-dotted lines. The uppermost phase without label is the
restored phase.
The letters denote the different phases,
where 
g: $\bar s s$,
l: $K^0 / \bar K^0$, 
n: 2SC,
q: 2SC + $\bar s s$ and 
t: CFL. 
}
\label{fig:muud550must}
\end{figure}

The authors of Ref.\ \cite{Barducci2005} used different quark masses
and a different coupling constant, and in addition employed a form
factor to mimic asymptotic freedom. Therefore, one may conclude that
the use of such a form factor does not affect the phase diagram
qualitatively. We would also like to point out that the phase diagram Fig.\
\ref{fig:muudmust0} cannot simply be obtained by a superposition of
phase diagrams obtained from a calculation with pseudoscalar
condensation, but without superconductivity (such as done in 
\cite{Barducci2005}), and
one with superconductivity, but without pseudoscalar condensation
(such as done in \cite{Gastineau2002}). 
Despite the fact that the two types of phases do
not coexist, there is nevertheless competition between them. 
Figure \ref{fig:muudmust0} shows that the $K^0 / \bar K^0$ (l)/(m)
and the $K^+ / K^-$ (j)/(k) phases are separated from the 2SC phase (q) by a
first-order transition. This remains the case at finite temperature as
is illustrated in Fig.\ \ref{fig:muud550must}. This figure displays
the phase diagram as a function of $\mu_s$ and temperature, for fixed
$\mu_u = \mu_d = 550\;\mathrm{MeV}$.

Returning to the discussion of Fig.\ \ref{fig:muudmust0}; the line
$\mu_u = \mu_d = \mu_s$ goes through the phase (a) in which chiral
symmetry is spontaneously broken. At some point it enters via a
first-order transition the 2SC + $\bar s s$ phase (q), and finally
goes into the CFL phase (t), again via a first-order transition. If
there is a difference between $\mu_u = \mu_d$ and $\mu_s$, one can see
from Fig.~\ref{fig:muudmust0} that as the densities increase, quark
matter can go directly from a phase of spontaneous chiral symmetry
breaking (a) to a CFL phase (t) without passing the 2SC phase (q) first. This
can also occur in compact stars \cite{Alford2002}. One should keep in
mind though that the relation between chemical potential and number density
is not linear. For example, at a first-order phase boundary, the
number density increases discontinuously. At these particular
densities, quark matter can be in a mixed state of normal and
superconducting matter \cite{Bedaque2002, Lawley2005}.

It is also interesting to note that the phases (l)/(j) of kaon
condensation can also occur outside the region of spontaneous chiral
symmetry breaking. Assuming the phase transition towards chiral
symmetry restoration coincides with the deconfinement transition (as
appears to be the case in lattice studies at small baryon chemical
potential and in some models), this would imply that condensation of a
state with quantum numbers of the kaon may persist in the deconfined
phase. This was first observed in Ref.~\cite{Son2001}, based on a perturbative
calculation at high isospin chemical potential, that is expected to be 
applicable only in the deconfined region. 
 
\subsection{$\mu_u = \mu_s$}

\begin{figure}[htb]
\includegraphics{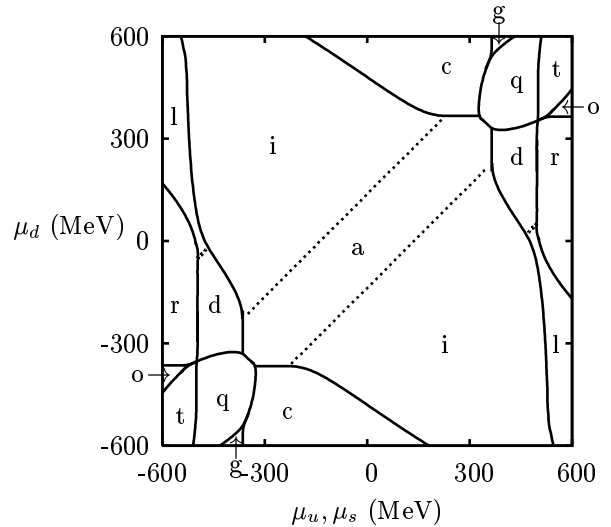}
\caption{Phase diagram for $T = 0$ as a function of $\mu_u=\mu_s$ and
$\mu_d$. First and second-order transitions are indicated
by solid and dotted lines, respectively. The letters denote
the different phases, where 
a: $\bar u u$ + $\bar d d$ + $\bar s s$,
c: $\bar u u$ + $\bar s s$, 
d: $\bar d d$ + $\bar s s$,
g: $\bar s s$, 
i: $\pi^+ / \pi^-$ + $\bar s s$ , 
l: $K^0 / \bar K^0$, 
o: 2SCus, 
q: 2SC + $\bar s s$,
r: 2SCus + $\bar d d$ and
t: CFL.}
\label{fig:muusmudt0} 
\end{figure}

In Fig.\ \ref{fig:muusmudt0}, we show the phase diagram at zero
temperature as a function of $\mu_u = \mu_s$ and $\mu_d$. This diagram
is similar to Fig.\ \ref{fig:muumudt0} for small strange quark
chemical potentials (below the kaon mass). 
At larger strange quark chemical potentials the
diagrams differ, exhibiting kaon condensation (l) and
diquark condensation involving strange quarks (r)/(o)/(t).

\begin{figure}[t]
\includegraphics{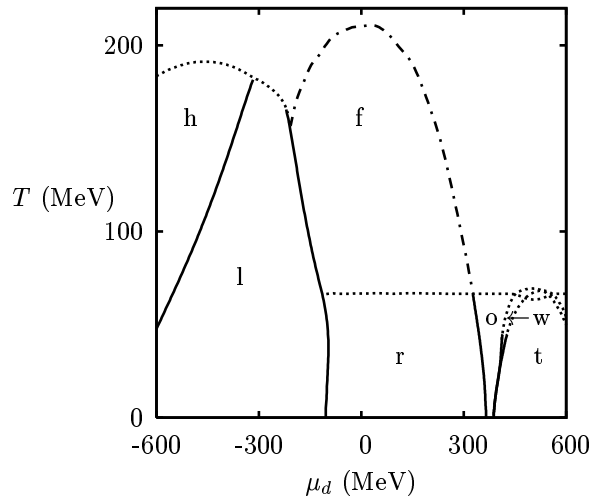}
\caption{Phase diagram as a function of $\mu_d$ and $T$, for fixed
$\mu_u = \mu_s = 550\;\mathrm{MeV}$. First and second-order 
transitions are indicated
by solid and dotted lines, respectively, while cross-overs
are denoted by dashed-dotted lines. The uppermost phase without label is the
restored phase.
The letters
denote the different phases, where f: $\bar d d$, h: $\pi^+ / \pi^-$,
l: $K^0 / \bar K^0$, o: 2SCus, r: 2SCus + $\bar d d$, t: CFL and w:
sSC. The lower right corner of this figure is enlarged in Fig.\
\ref{fig:muus550mudtr}.}
\label{fig:mus550mudt} 
\end{figure}

\begin{figure}[t]
\includegraphics{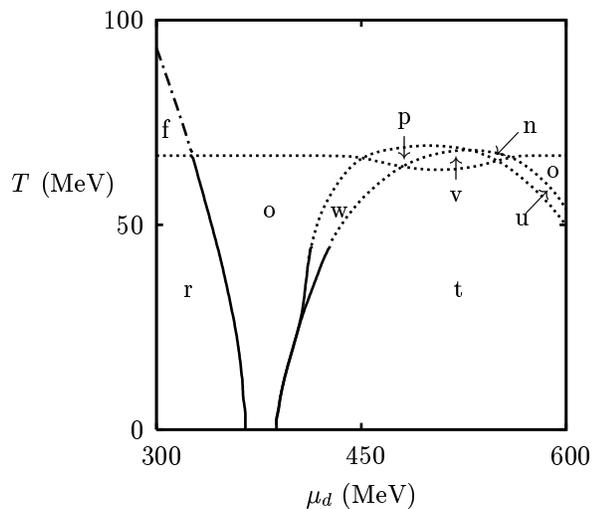}
\caption{Same as Fig.\ \ref{fig:mus550mudt}. The phases that occur are 
f: $\bar d d $, 
n: 2SC,
o: 2SCus, 
p: 2SCds,
r: 2SCus + $\bar d d$, 
t: CFL,
u: uSC,
v: dSC and
w: sSC.}
\label{fig:muus550mudtr} 
\end{figure}

In Fig.\ \ref{fig:mus550mudt}, we show the phase diagram as a function
of $\mu_d$ and $T$, for fixed $\mu_u = \mu_s = 550\;\mathrm{MeV}$. In
this figure one can find five critical points. Also, one can see in
this figure that the $K^0 / \bar K^0$ phase (l) is separated from the
2SCus phase (r) by a first-order transition for all temperatures.
Furthermore, it is interesting that at finite temperature there is a
first-order transition from the phase in which the neutral kaons
condense (l) to the pion condensed phase (h). In Fig.\
\ref{fig:muus550mudtr}, we have enlarged the lower-right corner of
Fig.\ \ref{fig:mus550mudt} for clarity. In this figure one can find
all possible superconducting phases, including the more exotic uSC (u), dSC (v) and 
sSC (w) phases. For $\mu_u = \mu_d = \mu_s$ one goes from the
CFL phase (t) via the 2SC phase (n) to the chirally restored
phase when raising the temperature. However, small differences 
between $\mu_u = \mu_s$ and $\mu_d$ can cause one to go through completely 
different phases.

\section{Conclusions}
In this paper, we studied the phase diagram of the three-flavor NJL
model including pseudoscalar condensation and color superconductivity
as a function of the different quark chemical potentials and
temperature. The NJL model has a rich and
interesting phase structure. The pseudoscalar condensed and color
superconducting phases are competing and are separated by a
first-order phase transition. As we have discussed, this need not be
the case for other (less conventional) choices of the parameters of
the model.

Furthermore, we concluded that at zero temperature and zero strange
quark chemical potential, there is a minimum asymmetry of about 35 MeV
between the up and the down quark chemical potentials required in
order to have two first-order transitions, when going from the phase
with spontaneous chiral symmetry breaking to the 2SC phase.

Our results provide a qualitative check and extension of several earlier
calculations that appeared in the literature. The new aspects of the
phase diagrams are often located in regions, where the quark chemical
potentials are large and very different in magnitude for the
different flavors.  Although such situations are not necessarily
realized in compact stars or can be realized in heavy-ion collisions,
a comparison with future lattice data may nevertheless provide
interesting information.  This is especially relevant for pseudoscalar
condensation in the phase where chiral symmetry is restored and also
for the complicated superconductivity phase structure close to the
cutoff of the model.

This work can be extended in several ways. For example, one can take
into account 't Hooft's instanton-induced interaction
\cite{tHooft1976}. If one has pseudoscalar condensation, this is more
difficult than in the normal case. Another useful extension would be the
inclusion of the neutrality conditions~\cite{Alford2002, Steiner2002},
in which case the phase structure
changes and for instance gapless phases will occur. 
It would also be interesting to
see how the results depend on the strength of the diquark coupling and
also on the quark masses. Furthermore, one could add the LOFF phase
\cite{Alford2001b}. In this crystalline phase, quarks of different
momenta can pair. One could also include vector interactions. In this
case spin-1 diquark condensation (see for example Refs.\
\cite{Pisarski1999, Buballa2002}) and an induced Lorentz-symmetry
breaking (ISB) phase~\cite{Langfeld1998} are among the
possibilities. It would also be worthwhile to take pseudoscalar diquark
condensation \cite{Buballa2004b} into account. Finally,
one could try to go beyond the mean-field approximation as was done in
Ref.\ \cite{Hufner1994}.

\section*{Acknowledgments}
The research of D.B.~has been made possible by financial support from
 the
Royal Netherlands Academy of Arts and Sciences.

\end{document}